\documentclass[fleqn,usenatbib]{mnras}
\usepackage{amsmath}
\usepackage{graphicx}
\usepackage{epstopdf}
\usepackage{amssymb,txfonts}
\usepackage{url}

\newcommand{\g}{$\gamma$}

\newcommand{\mbh}{M_{\rm BH}}
\newcommand{\msun}{{\rm M}_{\sun}}
\newcommand{\ledd}{L_{{\rm Edd}}}
\newcommand{\mdot}{\dot M}
\newcommand{\medd}{\dot M_{\rm Edd}}
\newcommand{\ergs}{\rm \,erg\,s^{-1}}

\newcommand{\kev}{\rm \,keV}
\newcommand{\lx}{L_{\rm X}}
\newcommand{\lr}{L_{\rm R}}
\newcommand{\chandra}{{\textit{Chandra}}}

\newcommand{\xte}{{\textit{RXTE}}}
\newcommand{\xmm}{{\textit{XMM--Newton}}}
\newcommand{\sax}{{\textit{Beppo\-SAX}}}

\newcommand{\nustar}{{\textit{NuSTAR}}}
\newcommand{\suzaku}{{\textit{Suzaku}}}
\newcommand{\fermi}{{\textit{Fermi}}}

\newcommand{\swift}{{\textit{Swift}}}

\title[An LHAF in NGC~7213]{A luminous hot accretion flow in the low-luminosity\\ active galactic nucleus NGC~7213}

\author[F. G. Xie et al.]{Fu-Guo Xie$^1$\thanks{E-mail: fgxie@shao.ac.cn (FGX); aaz@camk.edu.pl (AAZ)}, Andrzej A. Zdziarski$^2$$^\star$, Renyi Ma$^3$ and Qi-Xiang Yang$^1$ \\
$^1$Key Laboratory for Research in Galaxies and Cosmology, Shanghai Astronomical Observatory, \\
Chinese Academy of Sciences, 80 Nandan Road, Shanghai 200030, China\\
$^2$Nicolaus Copernicus Astronomical Center, Polish Academy of Sciences, Bartycka 18, PL-00-716 Warszawa, Poland\\
$^3$Department of Astronomy and Institute for Theoretical Physics and Astrophysics, Xiamen University, Xiamen, Fujian 361005, China}

\pubyear{2016}

\begin{document}
\label{firstpage}
\pagerange{\pageref{firstpage}--\pageref{lastpage}}
\maketitle

\begin{abstract}
The active galactic nucleus (AGN) NGC~7213 shows a complex correlation between the monochromatic radio luminosity $\lr$ and the 2--10 keV X-ray luminosity $\lx$, i.e. the correlation is unusually weak with $p\sim 0$ (in the form $\lr\propto\lx^p$) when $\lx$ is below a critical luminosity, and steep with $p>1$ when $\lx$ is above that luminosity. Such a hybrid correlation in individual AGNs is unexpected as it deviates from the Fundamental Plane of AGN activity. Interestingly, a similar correlation pattern is observed in the black hole X-ray binary H1743--322, where it has been modelled by switching between different modes of accretion. We propose that the flat $\lr$--$\lx$ correlation of NGC~7213 is due to the presence of a luminous hot accretion flow, an accretion model whose radiative efficiency is sensitive to the accretion rate. Given the low luminosity of the source, $\lx\sim 10^{-4}$ of the Eddington luminosity, the viscosity parameter is determined to be small, $\alpha\approx 0.01$. We also modelled the broad-band spectrum from radio to $\gamma$-rays, the time lag between the radio and X-ray light curves, and the implied size and the Lorentz factor of the radio jet. We predict that NGC~7213 will enter into a two-phase accretion regime when $\lx \ga 1.5 \times 10^{42}\, {\rm erg\,s^{-1}}$. When this happens, we predict a softening of the X-ray spectrum with the increasing flux and a steep radio/X-ray correlation.
\end{abstract}

\section{Introduction}
\label{sec:intro}

There is a general consensus that Seyferts, low-ionization nuclear emission-line regions (LINERs), radio galaxies and some transition galaxies contain supermassive black holes (BHs) accreting at moderately low accretion rates \citep{Ho08}. We call them here low-luminosity active galactic nuclei (LLAGNs). They have the bolometric luminosity of $L_{\rm bol}\la (0.01$--$0.02)\ \ledd$, where the Eddington luminosity (for the H fraction of 0.7) is $\ledd\simeq 1.47\times 10^{46} \ergs\ (\mbh/ 10^8 \msun)$ and $\mbh$ is the BH mass. LLAGNs have properties distinctly different from bright AGNs, see \citet{Ho08} for a review. In particular, the spectra of LLAGNs generally do not have big blue bumps, but instead peak in the mid-infrared \citep{Ho09}. Also, the optical--X-ray spectral index, $\alpha_{\rm OX}$ \citep{Sobolewska09,Sobolewska11}, the radio loudness \citep*{Ho02, Greene06} and the X-ray photon index, $\Gamma$ (\citealt{Gu09,Emm2012}, hereafter E12; \citealt{Yang2015, Connolly2016}), are all observed to anti-correlate with the Eddington ratio (reported either for the bolometric or X-ray band) in LLAGNs, while they correlate positively in bright AGNs.

These differences show that LLAGNs are not simply bright AGNs scaled down in the accretion rate. The leading theoretical picture of LLAGNs is the accretion--jet scenario. This model utilizes three components \citep*{Esin1997, YCN2005, Ho08, Yuan2014}, namely a cold geometrically-thin Shakura-Sunyaev disc (SSD; \citealt{SS1973}) truncated at certain transition radius, $R_{\rm tr}$, a hot accretion flow within this radius, and a jet. The hot component can either be an advection-dominated accretion flow (ADAF; \citealt{Narayan1994}) when the accretion rate, $\mdot$, is low or a luminous hot accretion flow (LHAF; \citealt{Yuan2001, Yuan2003}) when $\mdot$ is intermediate. The accretion-jet model has been successfully applied to both LLAGNs and the hard state of accreting BH binaries (BHBs), see \citet{Yuan2014} for a recent review.

We study here NGC~7213, a nearby face-on Seyfert 1/LINER with a number of interesting properties. The source redshift is $z_{\rm r}=0.005839$, and its BH mass is estimated as $\mbh=9.6^{+6.1}_{-4.1}\times 10^7\ \msun$ \citep*{Blank2005}. Using $H_0=67.8$ km s$^{-1}$ Mpc$^{-1}$ \citep{planck} and the redshift corrected to the reference frame defined by the microwave background\footnote{\url{http://ned.ipac.caltech.edu}} of 0.005145, the luminosity distance, $d_{\rm L}$, is 22.8 Mpc. The bolometric luminosity has been estimated as $L_{\rm bol} \simeq 9 \times 10^{42}$ and $\simeq 1.8 \times 10^{43}\ergs$ by \citet{SSNF2014} and E12, respectively, though the latter estimate may contain an infrared (IR) contribution from the local starburst. At the best-fitting mass, those values correspond to $\simeq 0.6\times 10^{-3}$ and $\simeq 1.3 \times 10^{-3}\,L_{\rm Edd}$, respectively.

The monochromatic radio (defined as $\lr=\nu L_\nu$ at, e.g., 5 or 8.5 GHz) luminosity, the 2--10 keV X-ray luminosity and the BH mass in AGNs and BHBs were found to be relatively tightly correlated \citep*{Merloni2003, Falcke2004}, forming the so-called Fundamental Plane (hereafter FP). The FP can be expressed as $\lr\propto \lx^p\ \mbh^q$ with the indices of $p=0.6\pm 0.1,\, q=0.78\pm 0.11$ \citep{Merloni2003}. The case of BHBs, where there is almost no dependence on the BH mass, has been reviewed by \citet{Corbel2013}. Physically, the FP provides evidence for a strong connection between the hot X-ray emitting source, and the radio source, usually a jet. In the accretion-jet model, the radio is from the jet, and unless the source is sufficiently faint the X-rays will emit from the hot accretion flow. The FP is interpreted as the consequence of a tight coupling between accretion and ejection, which is most evident in the mass flow ratio $\eta_{\rm jet}$ (cf. Sec. \ref{sec:rxmodel} for definition), see \citet{YC2005} and \citet{Xie2016} for details.

\citet{Bell2011}, herafter B11, carried out intense monitoring of NGC~7213 in radio and X-rays over three years. They found a significant correlation, but with a clear deviation in its slope with respect to the FP. Similar deviations are also observed in a few other AGNs (NGC~4051 and NGC~4395; \citealt{King2011, King2013}) and BHBs, where they have been named outliers \citep{Coriat2011, Corbel2013, Xie2016}. The latter seem to follow a hybrid correlation, most evident in H1743--322 \citep{Coriat2011}, which is the outlier with the largest dynamic range in $\lx$. The radio/X-ray correlation of H1743--322 exhibits three branches \citep{Coriat2011}: it is steep, $p\approx 1.3$, at large $\lx$, flat, $p\sim 0$, at moderate $\lx$, and has the original FP slope, $p\approx 0.6$, at low $\lx$.

\begin{figure*}
\centering
\includegraphics[width=13.cm]{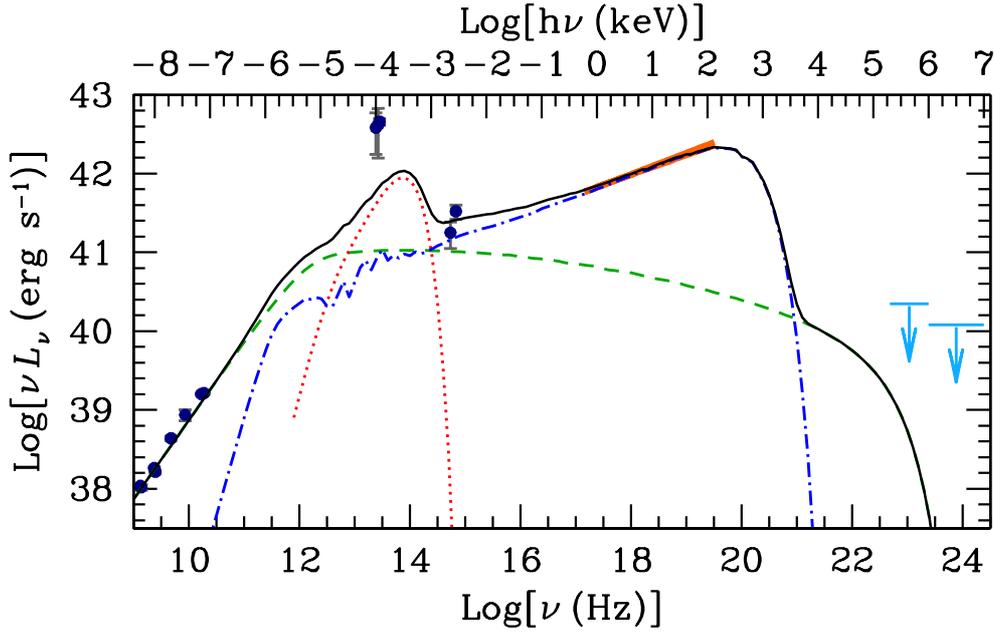}
\caption{The broad-band SED of NGC~7213. The red dotted, blue dot--dashed and green dashed curves show the emission from the truncated SSD, the LHAF and the jet, respectively. The black solid curve gives the total model. The radio to UV data are from E12, while the X-ray (0.6--150 keV) data, shown by the thick red line, are from \citet{Lobban2010}. The \fermi/LAT data are taken from \citet{Woja2015}.
}
\label{sed}
\end{figure*}

\begin{figure}
\centering
\includegraphics[width=8.cm]{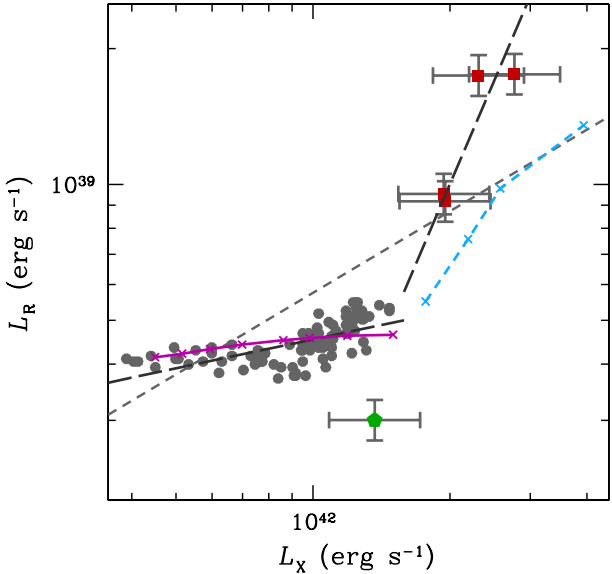}
\caption{The radio/X-ray correlation of NGC~7213, based on B11. The grey filled circles are for 4.8 GHz versus the 2--10 keV luminosity during MJD $\simeq$53800--54900 (2006 July--2009 September), which have been shifted by the 35-d lag of radio with respect to X-rays (as in fig. 7 of B11).  The green filled pentagon and red filled squares show archival data (MJD 47527--52057) collected by B11, for which the average interval between the radio (at 8.4 GHz) and X-ray measurements was $\simeq$1.5 yr. As B11, the errorbars of these archival data are estimated as the fraction rms scatter of the 3 yr ATCA/\xte\ data. The two black long-dashed curves represent the fitting of the grey circles and the red squares, respectively. Our model results are shown by the magenta crosses connected by solid line and the blue crosses connected by dashed line, see Section \ref{sec:rxmodel} for details. They approximately follow relation of $\lr\propto \lx^{0.15}$ and $\lr\propto \lx^{1.3--1.4}$, respectively. For comparison, the FP fit for $\mbh=10^8\, \msun$ \citep{Merloni2003} is shown by the grey dashed line.}
\label{rx_corr}
\end{figure}

Here, we point out that the radio/X-ray correlation in NGC~7213 resembles that of H1743--322. \citet{Xie2016} recently investigated both the FP-type and the hybrid radio/X-ray correlations in BHBs under the accretion-jet model. In their work, systems with the FP relationship were postulated to stay in the ADAF accretion regime, while systems with hybrid radio/X-ray correlation would enter different hot accretion modes. In the latter case, they postulated that an LHAF is responsible for the flat branch (with $p\sim0$), and a two-phase accretion flow (also called type II LHAF) is responsible for the steep ($p\sim 1.3$) branch. We develop these ideas here for NGC~7213, arguing that its central engine also enters the LHAF regime.

We note that alternative interpretations exist for the steep branch (at high $\lx$). \citet{MM2014}, \citet*{Cao2014} and \citet{Qiao2015} interpreted it in terms of the disc-corona model, while \citet*{Yu2015} proposed that it is due to a radiatively-cooled magnetized hot accretion flow. However, none of those models can explain the flat transition branch, which apparently represents a connection between the FP-type correlation and the steep one.

This paper is organized as follows. Section \ref{sec:obs} describes main observational properties of NGC~7213. Section \ref{sec:model} describes the LHAF-jet model, which is then applied to NGC~7213 in Section \ref{sec:application}. This allows us to understand its broad-band radio-to-$\gamma$-ray spectrum, time lags of the radio emission with respect to the X-rays, and constraints on the size of the radio source. In Section \ref{compare}, we compare NGC 7213 to other sources. Section \ref{sec:summary} gives our conclusions.

\section{Observational properties of NGC~7213}
\label{sec:obs}

\subsection{The broad-band spectrum}
\label{broadband}

B11 and E12 classified NGC~7213 as radio-intermediate, between radio-quiet and radio-loud AGNs. A representative broad-band spectral energy distribution (SED) of NGC 7213 is shown in Fig.\ \ref{sed}. Hereafter, both the observational and theoretical luminosities are defined as isotropic ones, i.e., $L\equiv 4\upi d_{\rm L}^2 F$, where $F$ is the observed flux. The radio-to-optical data are the averages of E12. The average 1.3--19 GHz spectrum, from the Australia Telescope Compact Array (ATCA), has a spectral index of $\alpha_{\rm R}\approx 0.1$ ($F_\nu \propto \nu^{\alpha_{\rm R}}$). In the radio-monitoring campaign, B11 found the individual values of $\alpha_{\rm R}$ between 4.8 and 8.4 GHz to be within $-0.4$ and 0.5. However, by taking into account the time lag between the two frequencies (see below), we estimate the intrinsic values of $\alpha_{\rm R}\approx 0.1$--0.2. The angular size at 8.4 GHz is constrained by high spatial-resolution Australian LBA observations to $\la$3 mas in diameter \citep{Blank2005}, or equivalently, the projected physical size at this frequency, $A_{8.4}$, is $A_{8.4} \la 0.33\, {\rm pc}\approx 6.8\times 10^4\ R_{\rm g}$, where $R_{\rm g}=G\mbh/c^2$ is the gravitational radius at the best-fitting black hole mass. The radio flux during that observation was about $\sim 40\%$ the flux shown in Fig.\ \ref{sed}, and we also adopt this size constraint during modelling.

The IR data from sub-arcsecond VLT/VISIR observations \citep{Honig2010, Asmus2011} shown in Fig.\ \ref{sed} probably include a contribution from the host galaxy. Based on high-spatial-resolution observations by Gemini, \citet{RD14} found a clumpy torus at the nucleus of NGC~7213. The viewing angle of this torus, and also that of the accretion flow if they are aligned, can be constrained through modelling of the IR spectrum, which yields $i\simeq 21\degr^{+9}_{-12}$ \citep{RD14}. The optical data shown are from the SMARTS/ANDICAM, after the subtraction of contribution from the host galaxy.

As typical for LLAGNs, we see that the UV bump is either absent or very weak in this source \citep{Starling2005,Lobban2010}. There is also no evidence for a Compton reflection continuum, and the observed narrow Fe K$\alpha$ line is probably produced in the broad-line region \citep{Bianchi2003, Bianchi2008, Starling2005, Lobban2010, Ursini2015}. The absence of the UV bump, relativistically broadened lines and reflection strongly suggests that a cold SSD in this source is truncated at a large radius (\citealt{Starling2005, Lobban2010}; E12).

The X-ray spectrum shown in Fig.\ \ref{sed} is from the joint \suzaku\/ + \swift/BAT observations \citep{Lobban2010}. The spectral index is well constrained, $\Gamma\simeq 1.74\pm 0.01$ ($F_E\propto E^{1-\Gamma}$), and there is no visible high-energy cutoff, with the e-folding cutoff energy $E_{\rm c}\ga 350 \kev$. The source has not been detected in the GeV energy band during 6.4 yr of \fermi/LAT observations \citep{Lobban2010, Woja2015}.

\subsection{The radio/X-ray correlation}
\label{sec:rxobs}

The radio/X-ray correlation in NGC~7213 has been studied by B11 during a 3 yr monitoring campaign by ATCA and \xte. We note that during their monitoring campaign, B11 also found time lags between different wavebands. They found that the emission at 8.4 and 4.8 GHz lag the X-rays by $\delta t_{8.4}=24\pm 12$ d and $\delta t_{4.8}=40\pm 13$ d, respectively. Also, the 4.8 GHz flux lags that at 8.4 GHz by $20.5\pm 12.9$ d. If consider the brightest and isolated flare of the campaign (MJD 53800--54000) alone, B11 derived $\delta t_{4.8}=35\pm 16$ d, which is more accurate compared to the 40-d lag in the sense that the peak value of the discrete cross-correlation function is larger (B11; Bell, private communication). The grey filled circles in Fig.\ \ref{rx_corr} show the data taken directly from fig.\ 7 in B11, where a 35 d time lag correction to the whole ATCA/\xte\/ data set is adopted. We see that the {\it averaged} radio and X-ray luminosities of NGC~7213 are roughly consistent with the FP (black dashed line), as noted by B11. However, the individual data clearly show a different correlation slope. The faint branch of the grey dashed curve shows a linear fitting between $\log\lr$ and $\log\lx$ to these ATCA/\xte\/ data points, which takes the form $\log (\lr/10^{39}\ergs) = (0.20\pm0.06) \log (\lx/10^{42}\ergs) -0.32\pm0.02$.

In Fig.\ \ref{rx_corr}, the green filled pentagon and red filled squares show the archival data (MJD 47527--52057) collected by B11, for which the average interval between the radio (at 8.4 GHz) and X-ray measurements was $\simeq$1.5 yr. Following B11, the errorbars of these archival data are estimated as the fraction rms scatter of the 3 yr ATCA/\xte\/ data. The jet remains unresolved at 8.4 GHz with increasing resolution to as high as 3 mas (\citealt{Blank2005}; B11 and references therein). Besides, the sensitivities of those observations with lower resolutions, i.e. the two red squares with $\lr\sim 9\times 10^{38}\ \ergs$ and those grey circles, are quite similar. The flux variability in radio is thus real (see also B11). A simple linear fitting between $\log\lr$ and $\log\lx$ to the red squares leads to $ \log (\lr/10^{39}\ergs) = (2.33\pm0.80) \log (\lx/10^{42}\ergs) -0.66\pm0.27$, as shown by the grey dashed curve (bright branch). In other words, we find that NGC~7213 is likely compatible to follow a steep, $p>1$, correlation when it is bright in X-rays. This result is admittedly less certain, given the possible existence of large-amplitude variability on time-scales shorter than the interval between the radio and X-ray observations. It is desirable to test this result in future, but we will tentatively assume its reality in this work.

Thus, NGC~7213 appears to feature a hybrid radio/X-ray correlation, similar to that in the BHB H1743--322 \citep{Coriat2011}. The maximal X-ray luminosity of the flat branch of the correlation branch is $L_{\rm X} \approx 1.5\times 10^{42}\ \ergs$ (corresponding to the 2--10 keV X-ray flux of $\sim 2.5\times 10^{-11}\, {\rm erg\,s^{-1}\,cm^{-2}}$). We will hereafter identify this maximum luminosity with the critical luminosity in the LHAF model, $L_{\rm X, LHAF, crit}$.

\subsection{The X-ray index--luminosity correlation}
\label{sec:globs}

Here, we compile X-ray spectral data for NGC 7213 from different detectors, which requires taking into account possible differences of their calibration. The main known difference appears to be that between the \xte/PCA and other instruments. In particular, \citet*{kdd14} have shown that the spectral indices from the PCA are systematically higher than those from the \xmm/EPIC-pn, by $\Delta\Gamma \simeq 0.1$. In order to correct for it, we subtract 0.1 from the values of $\Gamma$ for the used PCA observations.

\begin{figure}
\centering
\includegraphics[width=8.cm]{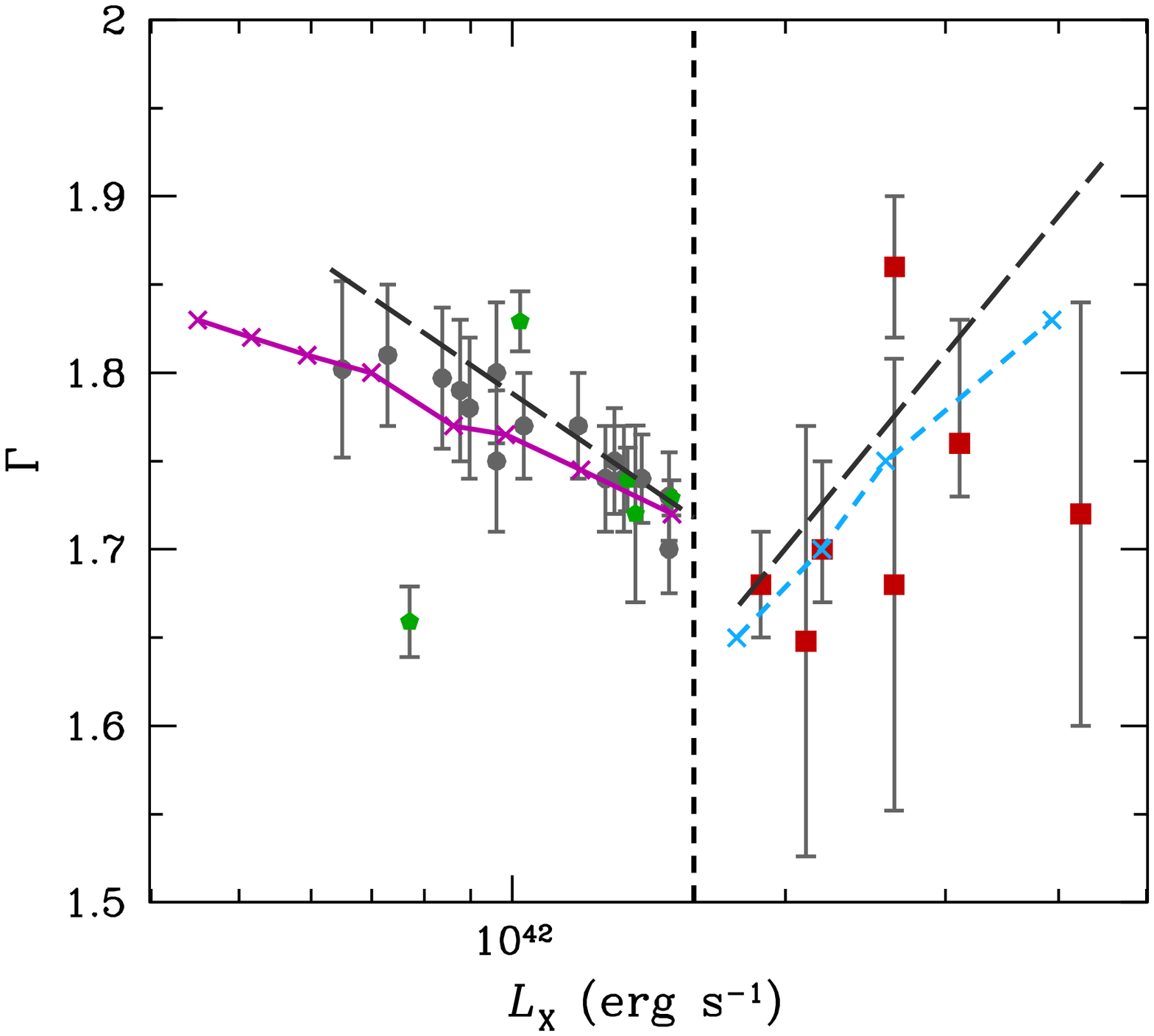}
\caption{The relationship between the X-ray photon index, $\Gamma$, and $\lx$. The grey filled circles are from the \xte\/ observations (E12, index corrected), where we see a negative correlation. The green filled pentagons show the results from \xmm\/ \citep{Emm2013}, \nustar\/ \citep{Ursini2015}, \xmm\/ \citep{Bianchi2003}, \chandra\/ \citep{Bianchi2008} and \suzaku\/ \citep{Lobban2010}, in the order of increasing $\lx$. Then, the red filled squares show the results from \textit{ASCA}\/ \citep{Turner2001}, \sax\/ \citep{Bianchi2003}, \textit{Ginga}\/ \citep{Nandra1994}, \sax\/ (lower, \citealt{Bianchi2003}), \textit{Ginga}\/ (upper, \citealt{Nandra1994}), \textit{EXOSAT}\/ \citep{Turner1989} and \textit{Einstein}\/ \citep{Halpern1984}, in the same order. The vertical dashed line marks the maximal LHAF luminosity in our model, $L_{\rm X, LHAF, crit}$. The two black long-dashed curves represent, respectively, the fitting of data above and below (green pentagon with lowest $\Gamma$ excluded from fitting) $L_{\rm X, LHAF, crit}$. The magenta crosses connected by solid line and blue crosses connected by dashed line represent our model, see Section \ref{sec:gammal} for details.
}
\label{gamma_L}
\end{figure}

The relationship between the X-ray index, $\Gamma$, and $\lx$ has been studied by E12 using the set of averaged PCA observations used by B11 for the radio/X-ray correlation except for its extension by an additional 100 d. Their results (with the corrected $\Gamma$) are shown by the grey filled circles in Fig.\ \ref{gamma_L}. We see a clear negative correlation. Such correlations are commonly observed in LLAGNs \citep{Connolly2016} and BHBs in their hard states,  see \citet{Yang2015} for a summary of observations and a theoretical interpretation.

We have also collected all other available X-ray data for this source, and we show their $\Gamma$-$\lx$ relationship in Fig.\ \ref{gamma_L}. We see that those other data below certain critical X-ray luminosity (green filled pentagons) are consistent with the presence of a negative correlation. We see here one outlier point, which may result from short-time-scale variability superimposed on the overall trend. On the other hand, there appears to be a positive correlation above that critical luminosity (red filled squares), though the error bars are relatively large, and more observations in the bright states are desirable. We note that the positive index--luminosity correlation is normally observed in bright AGNs (see \citealt{Z2003,SP2009,Yang2015} and references therein). We point out here that the critical X-ray luminosity constrained from the turnover in the V-shaped index--luminosity correlation is consistent with being the same as the value from radio/X-ray correlation, which we denoted as $L_{\rm X, LHAF, crit}$.

For completeness, we also do a linear fitting between $\Gamma$ and $\log \lx$, where the observational uncertainties in $\lx$ are also taken into account. As shown by the two black dashed curves in Fig.\ \ref{gamma_L}, the relationship is $\Gamma = (-0.35\pm0.06)\log(\lx/10^{42}\ergs) + 1.79\pm0.01$ for data with $\lx <L_{\rm X, LHAF, crit}$ (outlier excluded) and $\Gamma = (0.63\pm0.29)\log(\lx/10^{42}\ergs) + 1.51\pm0.12$ for data with $\lx > L_{\rm X, LHAF, crit}$.

\section{The LHAF-jet model}
\label{sec:model}

The LHAF-jet model is an extension of the ADAF-jet model to high accretion rates (\citealt{Esin1997,YCN2005}; see \citealt{Yuan2014} for a review on its dynamics, radiative properties, and applications). The ADAF has a critical accretion rate, $\propto\alpha^2$ \citep{Narayan1995, Xie2012}, where $\alpha$ is the viscosity parameter. Above this rate, the properties of the flow change, and it is called the luminous hot accretion flow (LHAF; \citealt{Yuan2001, Yuan2003}). From the energy equation of hot accretion flow (e.g. equation 4 in \citealt{Narayan1994}), it is found that the rate of radiative cooling, $q_{\rm rad}$, in LHAF is so strong that it exceeds the rate of viscous heating, $q_{\rm vis}$, in some regions of the hot accretion flow \citep{Yuan2001}. Consequently, the advection factor, defined as $f_{\rm adv} = q_{\rm adv}/q_{\rm vis}\equiv 1-q_{\rm rad}/q_{\rm vis}$ \citep{Narayan1994}, where $q_{\rm adv}$ is the entropy advection term, is negative in these regions. For comparison, the advection factor of ADAF, whose radiative cooling is dynamically unimportant, is always positive ($f_{\rm adv,ADAF}\sim 1 > 0$, cf. \citealt{Narayan1994, Yuan2014}). LHAF has been applied to model bright AGNs and BHBs in their bright hard and intermediate states (e.g., \citealt{Yuan2004, Yuan2007, Ma2012}).

Although LHAF is thermally unstable, the instability time-scale is longer than the dynamical time-scale if the accretion rate is below another critical accretion rate, corresponding to the 2--10 keV luminosity of $L_{\rm X, LHAF, crit}$. Consequently, the instability has no observational consequences \citep{Yuan2003}. However, above that rate, the thermal instability time-scale becomes shorter than the dynamical time-scale, and the cooling results in the formation of cold clouds mixed with the hot flow \citep{Yuan2003, Xie2012, Yang2015, Sadowski2016, Wu2016}. This two-phase accretion flow is also called type II LHAF, and the hot-phase-only LHAF above is renamed as type I LHAF (we will call it LHAF hereafter for simplicity, without introducing confusion). As the accretion rate increases even further, the entire flow will cool and and form a cold thin disc, with a hot corona above it \citep{Yang2015}. This is the usual disc-corona model, which is applied to bright AGNs and BHBs in their soft states \citep{Esin1997, Yuan2014}.

The LHAF shares the same dynamical equations and radiative processes with those of ADAF \citep{Yuan2001}. Strong outflows are found to exist in those flows (\citealt{Yuan2012b, Yuan2015}, and references therein). The accretion rate, $\mdot$, then is a function of radius, i.e. $\mdot(R)\propto R^s$ where $s<1$ is the outflow index (cf. \citealt{Yuan2012a} for the value of $s$ derived from numerical simulations). Here, we express $\mdot$ in units of the Eddington accretion rate, which we define as $\medd=10 \ledd/c^2$. Also, a fraction of the viscous power, $\delta$, is assumed to directly heat the electrons. The strength of the magnetic field in the flow is given by the ratio of the gas-to-magnetic pressures, $\beta\equiv p_{\rm gas}/p_B$. The dominant radiative cooling mechanism of LHAF is inverse Compton scattering of synchrotron photons.

The radiative efficiency, $L_{\rm bol}/\mdot c^2$, of hot accretion flows has been studied by \citet{Xie2012}, who found the efficiency of LHAFs to be very sensitive to the accretion rate, with $\lx \propto \mdot^{\sim 5-7}$. Such steep relation is mainly a consequence of strong inverse Compton at moderately optical depths. This feature is a key motivation for us to propose, following the work on the radio/X-ray correlations in BHBs \citep{Xie2016}, that NGC~7213 is in the (type I) LHAF regime when it shows a flat radio/X-ray correlation (cf.\ Section \ref{sec:rxmodel} for details). Furthermore, \citet{Xie2012} found that $L_{\rm X, LHAF, crit}$ is roughly proportional to $\alpha$. Thus, NGC~7213 could remain in the LHAF regime at luminosities as low as $\lx \sim 10^{-4} \ledd$ provided its value of $\alpha$ is small.

The two-phase accretion model is still far from being mature, because of its complicated radiative cooling processes with the existence of cold clumps. Currently, we are limited to a crude treatment of the radiative processes (see, e.g., \citealt{Xie2012, Yang2015}). One result of that model, which is of interest for this work, is that the radiative efficiency of two-phase accretion flow is nearly a constant, independent of $\mdot$ \citep{Xie2012}, i.e., $\lx \propto \mdot^{\sim 1}$, similar to that of SSD. Also, the two-phase accretion model produces a positive $\Gamma$--$\lx$ correlation \citep{Yang2015}.

Our jet model \citep{YCN2005} is based on the internal shock scenario, usually adopted in studies of \g-ray bursts. Compared to the original model, several improvements are adopted, including full consideration of relativistic effects and more accurate treatment on radiative transfer (see \citealt*{Z2012,Z2014}). The jet is assumed to consist of electrons and ions coming from the underlying hot accretion flow, and to be conical, with the half-opening angle of $\phi_{\rm jet}$. The bulk Lorentz factor is $\Gamma_{\rm jet}$ (corresponding to the velocity of $\beta_{\rm jet} c$) and the rate of the jet mass flow is $\mdot_{\rm jet}$. Collisions between shells with different velocities result in internal shocks, where a fraction, $\xi$, of the electrons is accelerated into a power-law distribution with the index, $p_{\rm e}$. High-energy electrons suffer strong radiative cooling, and their distribution steepens to $p_{\rm e}+1$, which has been taken into account self-consistently in the numerical calculations. In the relativistic shock acceleration theory, $2<p_{\rm e}<3$ is found, and we assume here $p_{\rm e}=2.2$ (e.g., \citealt{Kirk00, A01}). We also define in the comoving frame two dimensionless parameter, $\epsilon_{\rm e}$ and $\epsilon_B$, to measure the ratio of the energy density of power-law electrons and magnetic fields, respectively, to the shock energy density. The impacts of these parameters on the outcome spectrum of the jet are discussed in detail in \citet*{Xie2014} and \citet*{Qiao2015}.

\setlength{\tabcolsep}{3pt}
\begin{table}
\centering
\caption{LHAF-jet model parameters of NGC~7213 at $\lx\lesssim L_{\rm X, LHAF, crit}$}
\begin{tabular}{ccl}
\hline
Parameter & Value & Definition\\
\hline
\multicolumn{3}{c}{\it Truncated SSD}\\
\hline
$\mdot_0/\medd$ & $4.42\times 10^{-3}$ & $\mdot$ at $R_{\rm tr}$\\
$R_{\rm tr}/R_{\rm g}$ & $10^3$ & Truncation radius\\
\hline
\multicolumn{3}{c}{\it LHAF}\\
\hline
$s$ & $0.5$ & Outflow parameter\\
$\alpha$ & $0.011$ & Viscosity parameter\\
$\beta$ & $1.0$ & $p_{\rm gas}/p_B$\\
$\delta$ & $0.45$ & Fraction of electron viscous heating\\
\hline
\multicolumn{3}{c}{\it Jet}\\
\hline
$\mdot_{\rm jet}/\medd$ & $6.2\times 10^{-7}$ & Mass flow rate\\
$\phi_{\rm jet}$ & 0.1 & Half-opening angle\\
$\Gamma_{\rm jet}$ & $3$ & Bulk Lorentz factor\\
$\xi$ & 0.1 & Fraction of power-law electron\\
$p_{\rm e}$ & 2.2 & Electron power-law index\\
$\epsilon_{\rm e}$ & 0.02 & Acceleration-to-shock energy density ratio\\
$\epsilon_B$ & 0.02 & Magnetic-to-shock energy density ratio\\
\hline
\end{tabular}\label{parameters}
\end{table}

\section{Application to NGC~7213}
\label{sec:application}

\subsection{The spectral energy distribution}
\label{sec:sed}

\begin{figure}
\centering
\includegraphics[width=8.cm]{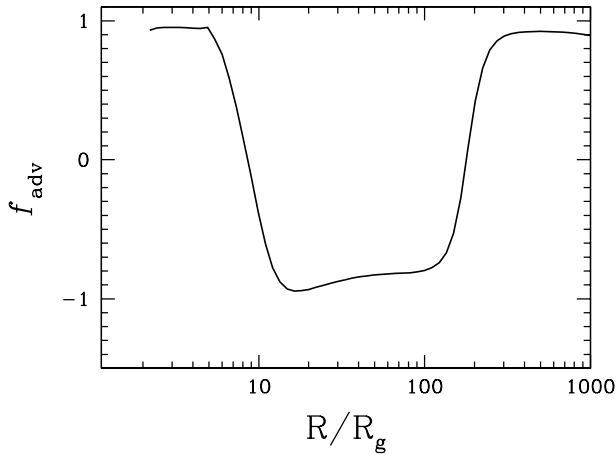}
\caption{The radial profile of the advection factor $f_{\rm adv}$ of the hot accretion flow derived during the SED modelling. We do observe a large region with $f_{\rm adv}<0$, confirming that it enters the LHAF regime.}
\label{fig:ngc7213fadv}
\end{figure}

We first apply the LHAF-jet model (see below and Fig.\ \ref{fig:ngc7213fadv} for confirmation on this accretion mode) to model the composite SED of NGC~7213. The SED, with $\lx\lesssim L_{\rm X, LHAF, crit}$, is shown in Fig.\ \ref{sed}. We fix $\mbh=10^8 \msun$ and $i=20\degr$, in agreement with the observational constraints. We have studied the parameter space of the LHAF-jet model applied to NGC~7213. The best model parameters are listed in Table \ref{parameters}. We note that the adopted magnetic field strength in LHAF is relatively high, given by $\beta=1$ (see also \citealt{Yang2015}). The strong magnetic field in LHAF, with $\beta\sim 1$, is supported by MHD simulations \citep{Machida2006} and numerical calculations \citep{Oda2009} confirming the expected effect of a decrease of the vertical scale height of the flow, $H$, with increasing cooling. This is likely to amplify the magnetic field energy density as the toroidal and radial magnetic fluxes are conserved, and thus $B^2\propto H^{-2}$. The crucial parameter of the LHAF-jet model is the obtained $\alpha=0.011$, which low value is required to keep the flow in the LHAF regime in spite of the low accretion rate. Fig.\ \ref{fig:ngc7213fadv} shows the profile of the advection factor, $f_{\rm adv}$, of the hot accretion flow component in the LHAF-jet model. We see that a large region of the hot flow has $f_{\rm adv} < 0$, i.e., it indeed enters into the LHAF regime (cf. Section \ref{sec:model} for differences between ADAF and LHAF), as expected. We also note that the jet bulk Lorentz factor, $\Gamma_{\rm jet}$, is degenerate with the jet mass flow rate, $\mdot_{\rm jet}$. Also, the jet parameters of $\xi$, $\epsilon_{\rm e}$ and $\epsilon_B$ have the values in their typical ranges.

The modelling results are shown in Fig.\ \ref{sed}, where the red dotted, blue dot--dashed and green dashed curves show the emission from the truncated SSD, the LHAF and the jet, respectively. The black solid curve represents the sum of all the three components. Since we do not include a torus during SED modelling, the model IR fluxes should be taken as lower limits. Except for those IR fluxes, the SED of NGC~7213 is well reproduced by our model. The radio emission is from the jet, while the UV up to soft $\gamma$-ray emission is from the LHAF. The high-energy cutoff in the theoretical spectrum is fitted as the e-folding energy of 620 keV, while the spectral index is $\Gamma=1.73$. These values are consistent with the joint \suzaku\/ + \swift/BAT data, see Section \ref{sec:obs}. We note that because the adopted accretion model is non-relativistic, we do not consider the $\gamma$-ray emission due to proton--proton interaction within the hot flow. This $\gamma$-ray emission is found to be very weak, consistent with current upper limit constraint, as examined recently by \citet{Woja2015}, where a general-relativistic version of hot accretion flow model is adopted.

\subsection{The radio/X-ray correlation: slope and normalization}
\label{sec:rxmodel}

We here follow \citet{Coriat2011} to discuss the relationship between radio and X-ray emission. In our jet model, the radio emission is due to partially self-absorbed synchrotron emission, which satisfies $\lr \propto \mdot_{\rm jet}^{\sim 1.4}$ \citep{Heinz2003}. If the X-ray emission from the hot accretion flow depends on the accretion rate as $\lx\propto \mdot^k$, and $\mdot_{\rm jet}\propto \mdot$ (though detailed numerical calculations yield a somewhat flatter dependence; \citealt{YC2005, Xie2016}), then we will have
\begin{equation}
\lr \propto \lx^{\sim 1.4/k}.\label{eq:rx}
\end{equation}

\citet{Xie2016} argued that the viscosity parameter in the FP-type, i.e., with $p\approx0.6$, sources is different from that in those hybrid sources. In accretion theory, the value of $\alpha$ determines the critical accretion rate (corresponding to some $\lx$) of different types of hot accretion flows. In the FP-type sources, the $\alpha$ is large, and the accretion flow remains in the ADAF (see also \citealt{YC2005}), which has $2<k<3$. In the hybrid sources like H1743--322, the value of $\alpha$ is considerably smaller. At low values of $\lx$, the accretion flow will be in the ADAF regime and we should observe an FP-type correlation. As $\lx$ increases, the hot accretion flow changes to the LHAF, whose $k\simeq 5$--7. In this regime, a small increase in $\mdot$ leads to a dramatic enhancement in $\lx$, as stated in Section \ref{sec:model}. At the same time, the increase of $\lr$ due to the small increase of $\mdot$ is also small. We thus observe a flat radio/X-ray correlation, cf. equation (\ref{eq:rx}) and note that $k\simeq 5$--7. As the $\lx$ increases further, the hot accretion flow enters the two-phase accretion flow regime, in which $k\simeq 1$ and a steep radio/X-ray correlation with $p\approx 1.3$--1.4 is observed.

Following the above arguments, we speculate that NGC~7213 enters into the (type I) LHAF regime for the flat radio/X-ray correlation branch, and the two-phase accretion (type II LHAF) regime for the steep branch. For the former regime, we find that a change of $\mdot_0$ by about 15 per cent results in a small change of $\lr$ but a change of $\lx$ by a factor of 3. For the latter regime, we find that we can reproduce a steep correlation with $p\sim 1.3$--1.4 (cf.\ \citealt{Coriat2011,Xie2016}). We show these results in Fig.\ \ref{rx_corr}, respectively, by the magenta and blue crosses. Here we assume the mass flow ratio between ejection and accretion, $\eta_{\rm jet} \equiv \mdot_{\rm jet}/\mdot (10R_{\rm g})$, to follow $\eta_{\rm jet} \propto \mdot (10R_{\rm g})^{\sim -0.5}$ in the LHAF regime and $\eta_{\rm jet} =$ const in the two-phase regime. These two relationships are motivated both theoretically \citep{Xie2016} and by the shape of the observed radio/X-ray correlation. Considering the uncertainties in observations, the flat branch is consistently reproduced. For the steep branch, our theoretical model also roughly reproduce the correlation slope, but the normalization is admittedly lower. This is because, the normalization of the $\eta_{\rm jet}$ profile is taken by the radio and X-ray data of the composite SED, where the radio is the {\it averaged} one of the whole ATCA observations. Obviously, the radio flux will be lower than the actual value at that given $\lx$, cf. the magenta cross with $\lx\lesssim L_{\rm X, LHAF, crit}$ and the corresponding grey circles.

Apart from the slope of the radio/X-ray correlation, the average radio and X-ray luminosities of NGC~7213 agree with the prediction of the FP, while the flat correlation branch in systems like H1743--322 lies well below the FP.  Possibly, this difference can be explained by the relativistic beaming effect \citep{Falcke1996} in NGC~7213, where the viewing angle is small, and which is classified as radio-intermediate (B11; E12).

\subsection{The index--luminosity correlation}
\label{sec:gammal}

As shown in Fig.\ \ref{gamma_L} and Section \ref{sec:globs}, NGC~7213 likely exhibits a V-shaped index--luminosity relation, i.e., its X-ray spectrum shows a harder-when-brighter behaviour when $\lx$ is low, and an opposite behaviour when $\lx$ is high. Below we follow \citet{Yang2015} to discuss the $\Gamma$--$\lx$ relationship. Theoretically the photon index is positively correlated with the ratio between the power received by the seed photons, $L_{\rm soft}$, and that received by the electrons, $L_{\rm hard}$ (see, e.g., \citealt{Z2003}). The key factor to determine the sign of the correlation is the origin of seed photons for Compton scattering. In the LHAF and ADAF regimes, the seed photons are dominated by the synchrotron emission. As $\mdot$ increases, a corresponding increase of $L_{\rm soft}$ is slow due to synchrotron absorption effect, and slower than the increase of $L_{\rm hard}$. Thus, we obtain a negative $\Gamma$--$\lx/\ledd$ correlation. On the other hand, the seed photons in two-phase accretion regime are quasi-thermally emitted by the cold clumps. As $\mdot$ increases, more and more clumps will be formed. Consequently, $L_{\rm soft}$ increases faster compared to $L_{\rm hard}$, and we obtain a positive $\Gamma$--$\lx/\ledd$ correlation.\footnote{\citet{Qiao2013} provide an alternative explanation under disc-corona model, where the seed photons of the positive $\Gamma$--$\lx$ branch are the thermal emission of inner cold disc.}

Detailed numerical results are shown in Fig.\ \ref{gamma_L}, where the magenta (connected by solid line) and blue (connected by dashed line) crosses correspond to the LHAF and two-phase accretion regimes. These theoretical predictions are in good agreement with the data.

\subsection{The radio--X-ray time lag and the size of radio emitting plasma}
\label{sec:timelag}

\begin{figure}
\centering
\includegraphics[width=8.cm]{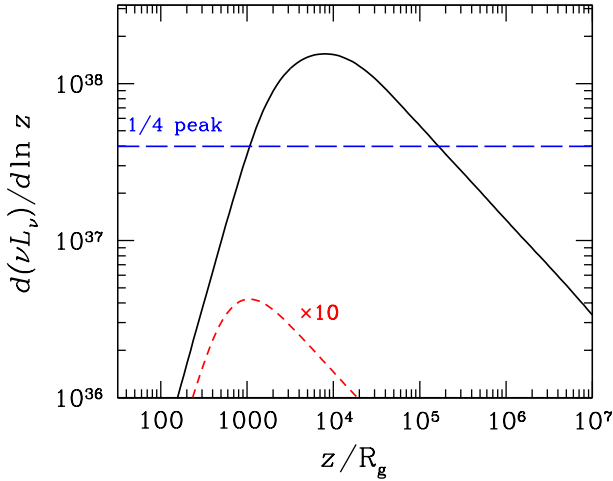}
\caption{The solid curve gives the profile of the (observed) radio emission at $8.4\ {\rm GHz}$ as a function of $z$. The dashed horizontal line intersects $1/4$ of the peak value, which is adopted to estimate the spread (in $z$) of the radio emission site. The emission from the counterjet, multiplied for clarity of display by a factor of 10, is shown by the red dashed curve.}
\label{fig:zloc}
\end{figure}

In the accretion-jet model, the bulk of X-rays come from a central region of the hot flow with the radius of $R\sim 10$--$40\ R_{\rm g}$, while the radio comes from the jet. A fluctuation in mass accretion rate $\mdot$ will have two consequences. First, it result in a flux change in X-rays. Secondly, the fluctuation will propagate into the jet, and will cause a change in $\mdot_{\rm jet}$. Thus, we will observe a correlated flux variability with changes in the radio range lagging those in X-rays. If the jet radio emission is partially self-absorbed, the bulk of emission at a given frequency comes from the distance where the emission becomes optically thin to synchrotron self-absorption \citep{Blandford1979}. In a conical jet, both the electron density and the magnetic field strength decrease with the height \citep{Blandford1979}. Thus, we also expect that emission at lower frequencies will lag those at higher ones. Both effects are confirmed by observations (Section \ref{sec:rxobs}).

In our model, the jet is launched from the surface of the hot accretion flow (in this case the LHAF) at $R\sim 10$--$20\ R_{\rm g}$. Because of the large aspect ratio $H/R$ of LHAF, we may safely argue that there is no time delay for part of the material in the dominate X-ray emitting region to enter into the jet. Consequently, the time lags between radio and X-rays, $\delta t_{8.4}$ and $\delta t_{4.8}$, allow us to measure the propagation time within the jet itself. For a simple conical jet with constant velocity, the bulk of emission at a given $\nu$ is occurs at the distance $z_\nu\propto \nu^{-1}$. We then have $\delta t_{4.8}/\delta t_{8.4} = z_{4.8}/z_{8.4}=8.4/4.8 = 1.75$. This is consistent with the observed value $\delta t_{4.8}/\delta t_{8.4} \simeq 1.7\pm 1.0$.

The time lag can be applied to constrain the jet bulk Lorentz factor. The observed, $\nu$, and emitted, $\nu'$, frequencies are connected by the Doppler factor, ${\cal D}\equiv 1/[\Gamma_{\rm jet}(1-\beta_{\rm jet}\cos i)]$, with $\nu'=(1+z_{\rm r})\nu/{\cal D}$. The distance from the BH of the bulk of the emission at a given $\nu$ is then approximately $\propto {\cal D}$. The observed time lag implies that the emission site at 8.4 GHz (observer frame) is at (see, e.g.,  \citealt{Ghisellini2000, Beckmann2012})
\begin{eqnarray}
z_{8.4} & = & \beta_{\rm jet} c \delta t_{8.4}/(1-\beta_{\rm jet}\cos i)\nonumber\\
& \approx & (4.2\pm 2.1) \times 10^3\ R_{\rm g}\ \beta_{\rm jet}/(1-\beta_{\rm jet}\cos i),\label{eq:z84}
\end{eqnarray}
where we assumed $\mbh=10^8\msun$.

Another independent constraint comes from the upper limit on the size of the plasma emitting at 8.4 GHz, $A_{8.4}$ (Section \ref{broadband}). We can write $A_{8.4}\approx\max (2 z_{8.4}\phi_{\rm jet}\cos i,\, \Delta z_{8.4} \sin i)$, where $\Delta z_{8.4}$ defines the spread of the emission site in the $z$-direction. Then the upper limit constraint in $A_{8.4}$ indicates that $z_{8.4}<3.6\times 10^5\ R_{\rm g}$ and $\Delta z_{8.4} < 2.0\times10^5\ R_{\rm g}$.

Together with the spectral modelling, we find that $\Gamma_{\rm jet}\la 4$ is required in order to obtain plausible jet parameters, where the upper limit of $\Delta z_{8.4}$ plays a key role. We set\footnote{Note that we assume a constant bulk velocity, while in reality the jet could be either accelerated or decelerated after the launch, as, e.g. in the M87 jet \citep{Asada2014}. However, we cannot constrain $\dot{\Gamma}_{\rm jet}$ based on the data available for NGC 7213.} $\Gamma_{\rm jet}=3$ ($\beta_{\rm jet}\simeq 0.94 c$, ${\cal D}\simeq 2.92$). Correspondingly, $z_{8.4}\approx (3.4\pm 1.7)\times 10^4 R_{\rm g}$ (cf.\ equation \ref{eq:z84}). In Fig.\ \ref{fig:zloc}, we show the emission distribution at 8.4 GHz, ${\rm d} (\nu L_\nu)/{\rm d}\ln z$, of both approaching (shown as the black solid curve) and receding (red dashed) jets, along the jet direction $z$. From this figure, we estimate $z_{8.4}\approx 1\times 10^4\ R_{\rm g}$. Besides, if we adopt $\sim$1/4 the peak emission to constrain the spread of the radio emission in the $z$-direction, then $\Delta z_{8.4}$ $\sim$ 1$\times 10^5\ R_{\rm g}$. Both values are consistent with current observations.

\section{Comparison to other objects}
\label{compare}

As discussed above, the radio/X-ray correlation in NGC~7213 appears to be similar to that observed in a number of BHBs classified as outliers. Also, the apparent change of the slope of $\Gamma$--$\lx/\ledd$ correlation in NGC~7213, from negative to positive, is similar to that found in many BHBs (e.g., \citealt{Yang2015} and references therein).

However, there appears to be no other AGN with such index--luminosity behaviour. This is probably because the evolutionary time-scale of AGNs are much longer than that of BHBs. Consequently, the dynamical range in $\lx/\ledd$ is usually much smaller than that in BHBs. An object showing some similarities is the bright AGN NGC~4051. Although this source mainly stays at the positive index--luminosity correlation branch (correspondingly the two-phase accretion regime in our model), \xmm\/ observations hint at a turnover at its faintest $\lx$, see the bottom panels of figs 9 and 10 in \citet{Ponti2006}. Thus, it may also enter into the (type I) LHAF regime there. However, the Very Large Array and \chandra\/ monitoring on this source at high X-ray fluxes, i.e. at the positive $\Gamma$--$\lx$ branch, shows a null (or possibly a weak negative) correlation between the radio and X-rays \citep{King2011}, which is inconsistent with the steep radio/X-ray correlation predicted by the two-phase model.

Thus, NGC~7213 remains a unique AGN so far. With its low $L_{\rm bol}\sim 10^{-3} \ledd$ and the radio/X-ray and index/$L_{\rm X}$ correlations with two branches each, it provides a unique opportunity to test the accretion theory.

\section{Conclusions}
\label{sec:summary}

Our main results are as follows.

We compile an exhaustive set of the observational properties of NGC~7213. These include its average broad-band spectrum, the relationships between its X-ray luminosity with the radio luminosity and the X-ray spectral index covering a broad range of $\lx$, and the radio properties.

We successfully model these properties through the accretion-jet model, which includes three components, i.e. a truncated cold disc, an inner hot flow and a jet. The inner hot flow changes its mode from the LHAF to two-phase accretion, which explains the changes of the slopes in the radio/X-ray and index/flux correlations. The critical 2--10 keV luminosity at which the change of accretion mode occurs is $L_{\rm X, LHAF, crit} \approx 1.5 \times10^{42} \ergs \approx 1.0\times 10^{-4} \ledd$. The model parameters of the composite SED with $\lx\lesssim L_{\rm X, LHAF, crit}$ are listed in Table 1. The most important parameter determined by us is the viscosity parameter, which is found to be relatively small. More explicitly, we have $\alpha\approx 0.011\ (\mbh/10^8\ \msun)^{-1}\ (d_{\rm L}/22.8\ {\rm Mpc})^2$, where the uncertainties in $\mbh$ and/or $d_{\rm L}$ are taken into account. This low $\alpha$ value allows the accretion flow to be in the LHAF regime at a relatively low bolometric luminosity. This determination is in agreement with \citet{Xie2016}, who proposed that systems with hybrid radio/X-ray correlation have small $\alpha$ values.

Based on the observed time lags, the radio flux and the radio size constraint, we find the jet Lorentz factor to be small, $\Gamma\la 4$. The location of the radio source at the observed $\nu=8.4$ GHz is $\sim 10^4 R_{\rm g}$.

\section*{Acknowledgements}
We thank Professor Feng Yuan for suggestions and comments, Dr  Dimitris Emmanoulopoulos for providing us with X-ray data on NGC~7213, Dr Erlin Qiao for a discussion of the X-ray properties of NGC~4051, Drs Shanshan Weng and Zhen Yan for a discussion on X-ray detectors, and Dr Martin Bell for detailed information on their radio/X-ray correlation result presented in B11. We appreciate the referee for careful reading and a detailed report. FGX was supported in part by the National Key Research and Development Program of China (2016YFA0400800) and the Youth Innovation Promotion Association of CAS (id.\ 2016243). FGX, RM and QXY were supported in part by the Natural Science Foundation of China (grants 11273042, 11333004, 11573051, 11633006, U1231106 and U153110128), the National Basic Research Program of China (973 Program, grant 2014CB845800), the Strategic Priority Research Program `The Emergence of Cosmological Structures' of CAS (grant XDB09000000), and the CAS/SAFEA International Partnership Program for Creative Research Teams. AAZ has been supported in part by the Polish National Science Centre grants 2012/04/M/ST9/00780, 2013/10/M/ST9/00729 and 2015/18/A/ST9/00746.

\label{lastpage}
\end{document}